# Efficient Micro-Mobility using Intra-domain Multicast-based Mechanisms (M&M)

Ahmed Helmy, Muhammad Jaseemuddin, Ganesha Bhaskara[*]

*Abstract*

One of the most important metrics in the design of IP mobility protocols is the *handover* performance. Handover occurs when a mobile node changes its network point-of-attachment from one access router to another. If not performed efficiently, handover delays, jitters and packet loss directly impact and disrupt applications and services. With the Internet growth and heterogeneity, it becomes crucial to design efficient handover protocols that are *scalable, robust and incrementally deployable*. The current Mobile IP (MIP) standard has been shown to exhibit poor handover performance. Most other work attempts to modify MIP to *slightly improve* its efficiency, while others propose complex techniques to *replace* MIP.

Rather than taking these approaches, we instead propose a *new architecture* for providing efficient and smooth handover, while being able to *co-exist* and inter-operate with other technologies. Specifically, we propose *an intra-domain multicast-based* mobility architecture, where a visiting mobile is assigned a multicast address to use while moving within a domain. Efficient handover is achieved using *standard multicast join/prune* mechanisms.

Two approaches are proposed and contrasted. The first introduces the concept *proxy-based mobility,* while the other uses *algorithmic mapping* to obtain the multicast address of visiting mobiles. We show that the algorithmic mapping approach has several advantages over the proxy approach, and provide mechanisms to support it.

Network simulation (using NS-2) is used to evaluate our scheme and compare it to other routing-based micro-mobility schemes - CIP and HAWAII. The proactive handover results show that both M&M and CIP shows low handoff delay and packet reordering depth as compared to HAWAII. The reason for M&M's comparable performance with CIP is that both use bi-cast in proactive handover. The M&M, however, handles multiple border routers in a domain, where CIP fails. We also provide a handover algorithm leveraging the proactive path setup capability of M&M, which is expected to outperform CIP in case of reactive handover.

## I. INTRODUCTION

The growth of mobile communications necessitates efficient support for IP mobility. IP mobility addresses the problem of changing the network point-of-attachment transparently during movement. When the mobile node moves away from its current network point-of-attachment, *handover* is invoked to choose another suitable point-of-attachment. In such an environment, handover latency and mobility dynamics pose a challenge for the provision of efficient handover.

Several studies [1][8] show that Mobile IP (MIP)[3], the proposed standard, has several drawbacks ranging from triangle routing and its effect on network overhead and end-to-end delays, to poor performance during handover due to communication overhead with the home agent. Several micro-mobility approaches attempt to modify some mechanisms in Mobile IP to improve its performance[4][5]. However, as we will show, such approaches suffer from added complexity and, in general do not achieve the best handover performance.

We follow a different approach to IP mobility using *multicast-based mobility (M&M)*. In such architecture, each mobile node is assigned a multicast address to which it joins through the access routers it visits during its movement. Handover is performed through standard IP-multicast join/prune mechanisms. Such approach, however, is not suitable for inter-domain IP mobility, for several reasons. First, the architecture requires *ubiquitous multicast deployment*, which is only partially supported in today's Internet. M&M should be designed for *incremental deployment,* and to allow co-existence with other IP mobility protocols. Second, the *multicast state* kept in the routers grows as the number of mobile nodes becomes larger. This problem may be alleviated using *state aggregation* techniques[38]. Third, allocating a globally unique multicast address for every mobile node requires a global *multicast address allocation* scheme, and wastes multicast resources. Furthermore, mobile nodes incur *security* delay with every handover, which may overshadow architectural mechanisms that attempt to reduce handover delays.

To alleviate these problems, we propose new schemes for *intra-domain* multicast-based micro-mobility that allow for incremental deployment. In this architecture, a mobile node is assigned a multicast address within a domain for use with *micro mobility*. The allocated multicast address is *locally scoped* (i.e., unique only domain-wide). This allows for domain-wide address allocation schemes. Packets are *multicast-tunneled* to the mobile node within the domain. The multicast address of a mobile does not change throughout its movement within the domain. This allows for lighter-weight security during handover, as it is used for micro-mobility (i.e., intra-domain).

[*] A. Helmy and G. Bhaskara, Electrical Engineering Department, University of Southern California. Email: {helmy,bhaskara}@usc.edu. M. Jaseemuddin, Electrical and Computer Engineering, Ryerson University. Email: jaseem@ee.ryerson.ca.

In this paper we present two different approaches to multicast-based micro mobility, one approach is based on *mobility proxies* and the other based on a novel scheme for *algorithmic mapping*. We compare such approaches and show that algorithmic mapping provides a more scalable and robust approach, and we develop efficient, yet simple, mechanisms to realize it. Furthermore, we conduct extensive simulations to compare the handover performance of our approach to other routing-based micro-mobility schemes. The proactive handover performance results show that our scheme performs as well as CIP and much better than HAWAII. Furthermore, it handles multiple border routers in a domain where CIP fails. The rest of the document is outlined as follows. Section II introduces multicast-based mobility. Section III provides overview of the intra-domain architecture, and discusses the proxy-based approach. Section IV describes the algorithmic mapping approach in detail. Section V gives evaluation and comparison results. Section VI discusses related work. We conclude in Section VII.

## II. Multicast-based Mobility (M&M)

Performance during *handover* is a significant factor in evaluating performance of wireless networks. IP-multicast[25][2] provides efficient location-independent packet delivery. The receiver-initiated approach for IP-multicast enables receivers to join to a nearby branch of an already established multicast tree. Multicast-based mobility (M&M)[1][8] uses this concept to reduce latency and packet loss during handover.

In multicast-based mobility, each mobile node (MN) is assigned a multicast address. The MN, throughout its movement, joins this multicast address through locations it visits. Correspondent nodes (CN) wishing to send to the MN send their packets to its *multicast* address, instead of unicast. Because the movement will be to a geographical vicinity, it is highly likely that the join from the new location, to which the mobile recently moved, will traverse a small number of hops to reach the already-established multicast distribution tree. Hence, performance during handover improves considerably. An overview of this architecture is given in Figure 1. As the MN moves, it joins to the assigned multicast address through the new access router (AR). Once the MN starts receiving packets through the new location, it sends a prune message to the old AR to stop the flow of the packets down that path. Thus completing the smooth handover process. In spite of its promise, we believe that many issues need to be addressed to realize multicast-based mobility in today's Internet. These issues include scalability, multicast address allocation, multicast deployment and security.

*Scalability of Multicast State*: The state created in the routers en-route from the MN to the CN is source-group (*S, G*) state. With the growth in number of mobile nodes, and subsequently, number of groups (*G*), the number of states kept in the routers increases. In general, if there are '*x*' MNs, each communicating with '*y*' CNs on average, with an average path length of '*l*' hops, then number of states kept in the routers is '*x*y*l*' states. Clearly, this *does not scale*.

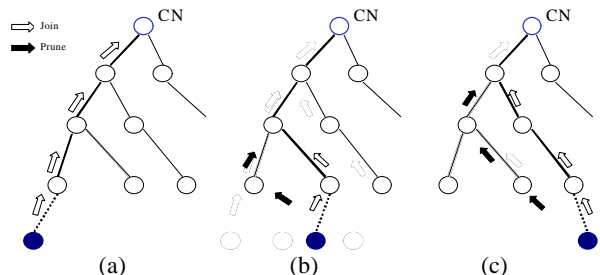

Figure 1: Multicast-based mobility. As the MN moves, as in (b) and (c), the MN joins the distribution tree through the new location and prunes through the old location.

*Multicast Address Allocation*: Inter-domain M&M requires each MN to be assigned a *globally unique* multicast address. Using a global multicast address for each MN may be wasteful and requiring uniqueness may not be practical[1].

*Ubiquitous Multicast Deployment*: Inter-domain M&M assumes the existence of inter-domain multicast routing. We believe, however, that incremental deployment and interoperability should be an integral part of any architecture for IP mobility.

*Security Overhead*: Security is critical for mobility support, where continuous movement of mobiles is part of the normal operation. Such setting is prone to *remote redirection* attacks, where a malicious node redirects to itself packets that were originally destined to the mobile. The problem is even more complex with multicast, where any node may join the multicast address as per the IP-multicast host model. These security measures are complex and may incur a lot of overhead. If such measures are invoked with every handover, however, it may overshadow the benefits of efficient handover mechanisms[2].

---
[1] Multicast address allocation is an active area of research [15]. We envision the number of MNs to grow tremendously.

To address the above issues, we propose a new approach for intra-domain multicast-based mobility.

## III. Intra-domain Architectural Overview

In our intra-domain architecture, a mobile node is assigned a multicast address to which it joins while moving. The multicast address, however, is assigned only within a domain and is used for *micro* mobility. While moving between domains, an inter-domain mobility (e.g., Mobile IP) protocol is invoked. In Mobile IP (MIP)[3], every mobile node (MN) is assigned a *home* address and *home agent* (HA) in its home subnet. When the MN moves to a *foreign* subnet, it acquires a care-of-address (COA) through a foreign agent (FA). The MN informs the HA of its COA through a *registration* process. Packets destined to the MN's home address are intercepted by the HA in the home subnet, then it *tunnels* them to the MN's COA. This is known as *triangle routing*. We will use the Mobile IP model to discuss inter-domain routing in the following sections.

Several mechanistic building blocks are needed to realize our proposed architecture. First, when the mobile moves into a new domain it is assigned a multicast address. What is the address allocation scheme? Second, packets destined to the mobile are multicast-tunneled by an *encapsulator* to the mobile node. How are the *encapsulator(s)* selected and where are they placed? To answer these questions, we investigate and compare two different approaches: (1) A *proxy*-based architecture, and (2) *Algorithmic mapping* architecture.

### A. Reference Architecture

We consider an IP network for a single domain, as shown in Figure 2. The network is connected to the Internet through Border Routers (BRs). An Access Point (AP) is the radio point of contact for a mobile node. A number of APs are connected to an Access Router (AR). From the access router's point of view, each AP is a node on a separate subnet. When a mobile moves from one AP to another without changing AR is an intra-AR handover case that can be specific to AR implementation and is not considered in this paper.

When a mobile moves into a new domain it is assigned a multicast care of address (MCOA). It is also assigned a unicast address that is unique within the domain, called regional care of address (RCOA). Since MCOA is used for routing packets within the domain, there is no need to assign COA at every subnet. The RCOA is a unique unicast address on the *m-subnet*. The m-subnet is a unique subnet that is characterized by the mobility where mobile nodes can use their RCOA to establish communication through any AR at the edge of the network. Hence, the m-subnet can be viewed as a logical subnet formed by all APs at the edge of the network. All ARs include the prefix for m-subnet in their router advertisements[37].

Address allocation and management is discussed later in this paper. When a mobile moves from one AR to another, it is said to handover from old AR ($AR_{old}$) to new AR ($AR_{new}$). We use this terminology throughout the rest of the paper.

First, we shall describe the proxy-based approach and discuss the problems associated with it.

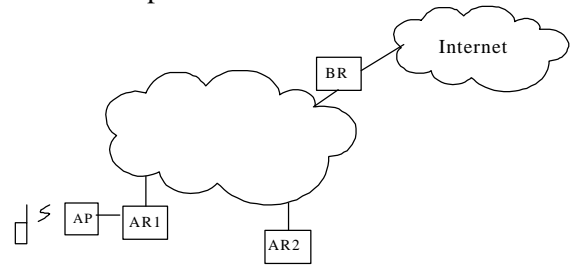

**Figure 2:** Reference Mobility Domain Network

### B. Proxy-based Architecture

When a mobile node moves into a new domain, it contacts its access router (AR). The AR performs the necessary per-domain authentication and security measures, and then assigns RCOA for the mobile node (MN). As shown in **Figure 3**, the AR then sends a *request* message to the mobility proxy (MP) to obtain a multicast address for the visiting MN. The request message includes the home address of the mobile node and its home agent's address. Upon receiving the request the MP performs two tasks. The first is to register on behalf of the mobile node its own address as

---

[2] Providing a comprehensive security solution for IP mobility is beyond the scope of this work. We believe, however, that our schemes relaxes security requirements during handover.

COA with the MN's home agent. The second task is to assign a multicast address for the visiting MN, send a *reply* message to the AR and keeps record of this mapping. The mapping is used for packet encapsulation later on. In this scheme, the MP remains transparent to the MN, which makes the placement of MPs within the domain flexible without notifying every MN.

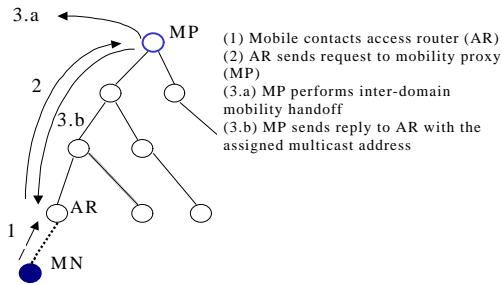

**Figure 3:** Event sequence as the mobile node moves into a domain.

Once this step is complete, the visiting MN joins the assigned multicast address ($G$). The joins are sent to the proxy-group pair ($MP, G$) and are processed as per the underlying multicast routing. The MN continues to move within the domain using the same multicast address. The scope of the assigned multicast address is local to the domain. Handover is performed using standard join/prune mechanisms and only lightweight intra-domain security is required in this case.

The HA tunnels the packets that are sent to the MN's home address to the MP using inter-domain mobility. The packets are then encapsulated by the MP, based on the mapping, and sent down the multicast tree to the MN. The MN uses the unicast RCOA for sending packets. To avoid single-point-of-failure scenarios multiple MPs are used. These MPs are typically placed at the border of the domain or at the center of the network[3]. An algorithm similar to[24] may be used for dynamic MP liveness and election mechanisms.

Several issues need to be addressed in the above architecture. First, the MPs need to maintain unicast-to-multicast address mapping for all visiting MNs. The scalability of such a scheme is of question. Second, complex robustness algorithms are needed to maintain MP liveness information, requiring initial configuration and setup. Third, the service disruption effect of MP failure is not clear. Since the MP registers its own address with the home agent and is used to encapsulate incoming packets, this introduces a *third-party-dependence* problem that is undesirable. In addition, MPs should run a multicast address allocation scheme to ensure collision-free address assignment.

To address these problems we propose a novel approach based on *algorithmic mapping* that obviates the need for explicit unicast-to-multicast mapping, and eliminates the need for complex address allocation.

### IV. Algorithmic Mapping Architecture

We provide mechanisms for address management and duplicate address detection, and inter-AR handover.

#### A. Overview

In this scheme we assume there is a one-to-one mapping between an RCOA and MCOA. When a mobile moves into a new domain it is assigned RCOA by the AR and the mobile performs *inter-domain handover* i.e., it registers the RCOA with its home agent. The AR *automatically infers* the multicast address (*MCOA*) for the mobile node from the assigned unicast address (RCOA) through a straight forward *algorithmic mapping*, described later in this section. The AR then triggers a Join message for MCOA to establish the multicast tree. Packets destined to the MN's home address are tunneled to its RCOA by the HA. These packets when arrive in the foreign domain are identified by the border router (BR) as being destined to a node on the m-subnet. As shown in Figure 4, the BR maps the destination unicast address to the *multicast address* and transmits the packets to the MN down the multicast tree. The serving AR changes the destination address from multicast to the unicast address. Since the destination address is modified twice within the network and restored to the RCOA by the AR, the packet does not cause security association violation at the mobile node.

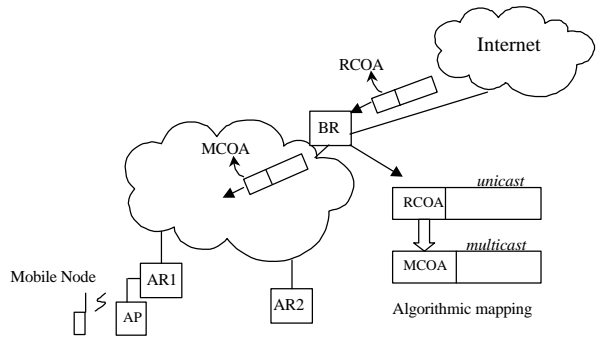

**Figure 4:** High level architectural view: Data packet is unicast over the Internet destined to the RCOA and arrives at the border router (BR) for the mobile node. The BR intercepts the packet and

---

[3] Network center are nodes with min(max distance) to any other node[26].

performs algorithmic mapping from the RCOA to MCOA. The packet is then multicast within the domain.

This architecture provides several *advantages* over the proxy-based approach. It *avoids the third party dependence* on the MP. Moreover, since algorithmic mapping is used, *no explicit RCOA-MCOA mapping is kept* or maintained by the encapsulator, which solves the mapping scalability problem and provides a more robust mechanism.

*B. Address Management*

The number of multicast addresses required is proportional to the number of mobile nodes in the domain. The scope of an MCOA is local to the domain where it is used. The IPv6 multicast addressing provides facility to define scope within the address[32]. Hence, in the rest of the paper we consider IPv6 address for both RCOA and MCOA.

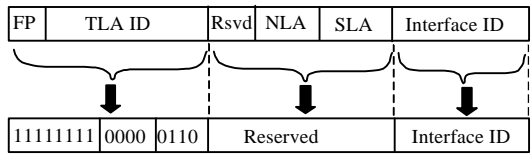

**Figure 5:** Algorithmic mapping

The standard IPv6 unicast and multicast address architectures[32] are shown in Figure 4 (a) and (b). We modify the group bits to include interface ID as the group ID. The remaining bits of the group ID is reserved that is ignored by multicast routing. The 64-bit interface ID address space is large enough for all the mobiles within a domain. We also define a new scope: micro-mobility scope with value 0x6. The SLA is a 16-bit long field, used to create local hierarchy and identify subnets[33]. A single subnet ID, identifying m-subnet, is defined for assigning RCOA.

When a mobile moves into a foreign domain it is assigned an RCOA. The AR forms the MCOA by replacing the <FP, TLA ID> bits of the RCOA with the multicast <FP, flag (0000), scope (0110)> values. This provides a simple, yet very efficient and unique *algorithmic mapping*. The mobile acquires RCOA on the m-subnet through either autoconfiguration or DHCP[34]. The auto-configuration requires duplicate address detection (DAD) [35] on every subnet. In our scheme the mobile obtains RCOA and MCOA once it is connected to the network. We propose a scheme in [39] that detects address duplication within the m-subnet, which is performed once at the AR during initial address assignment. The mobile afterward is able to move freely without running DAD at any other AR. When a mobile first connects to the network it performs a high latency inter-domain handover, hence duplicate address resolution latency is overshadowed by this handover latency.

*C. Intra-domain Handover*

When a mobile moves from one AR to another, a handover event takes place between the two routers. The handover involves *route repair* that is path setup inside the network to redirect the incoming traffic flow to the new AR. In *proactive handover* the link between the MN and new AR is established prior to its disconnection with the old AR. Hence a smooth handover, i.e. handover with low packet loss, can take place by exploiting the fact that the new AR is known a priori and bi-casting packets to both access routers. In *reactive handover* an abrupt disconnection may cause the MN to switch over to the new AR. The route repair in this case can only be initiated from the new AR, hence bi-casting cannot reduce packet loss. Multicasting allows *proactive path setup* to the new access router before the mobile is actually connected to it. This can minimize packet losses in reactive handover where bi-casting fails. Moreover, bi-casting being a special case of multicasting, multicasting-based solution, e.g. M&M, performs equally well for achieving proactive handover. In this section we describe one handover scheme where proactive path setup is used to achieve smooth handover.

We define a set of adjacent access routers as the Coverage Access Router Set (CAR-set). The adjacency can be established based on the adjacency of the radio coverage area of the serving AR in case of cellular wireless network. The serving AR is called the Head of the CAR-set. Thus, there is a unique CAR-set defined for every AR. For example, in Figure 6 AR1 to AR7 constitute a CAR-set for AR1, which is the serving AR for the mobile. The mobile can move to any of the ARs in the CAR-set without interruption in the packet flow.

A site-local multicast group address is assigned to each CAR-set, called CAR-set group address (CGA). Every AR that is a member of a CAR-set must join the corresponding CGA, which serves as a control channel for the members to exchange the control signals. For example, in Figure 6, all the access routers surrounding AR1 join CGA1 to become members of AR1's CAR-set (CGA1). Similarly, AR1 must also join six other CAR-sets corresponding to adjacent routers AR2 to AR7.

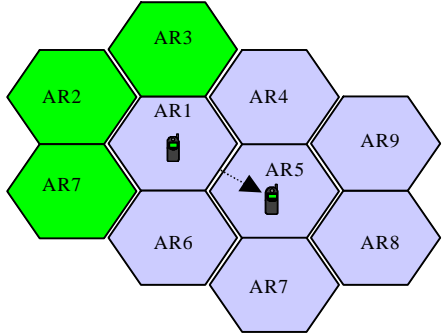

**Figure 6:** Handover across CARS

We define three new control signals as follows:
1. *J-message* causes the receiving router to *join* the multicast group identified in the message.
2. *L-message* causes the receiving router to *leave* the multicast group identified in the message.
3. *HO* message exchanged between the two routers involved in *handover*. Its parameter includes the mobile's RCOA and MCOA.

We explain the handover algorithm by using the example depicted in Figure 6. Consider the MN moving from AR1 to AR5. When connectivity is established between the MN and AR5, the AR5 multicasts a J-message <MCOA> to the members of its CAR-set (CGA5) requesting them to join the mobile's MCOA. It then sends HO <RCOA, MCOA> message to AR1 to initiate the prune process. When AR1 receives HO message it multicasts an L-message <MCOA> to members of its CAR-set (CGA1) requesting them to leave the MCOA.

Although the ordering of (J => HO => L) messages ensures that L-message is initiated after J. The order of message reception, however, is not guaranteed to both CAR-sets. Depending on the order of arrival of J and L messages at an AR that is a member of both CAR-sets, it may leave the MCOA whereas it is supposed to have remained joined to that group. To ensure consistency between Join and Leave messages we introduce the following mechanism. Each AR keeps its membership status in a 4-tuple <MCOA, Serving Access Router (SR), CGA, State> table. The table contains an entry corresponding to every mobile roaming in a CAR-set of which the access router is a member. There are two states defined: *Joined* and *Left*. The rules for updating the table specify that an AR only accept L-message for a MCOA, if the source of the L-message matches the SR in the MCOA's entry (i.e., the AR has joined the MCOA on the request of the same SR)[4]. Otherwise, the L-message is discarded. The AR accepts all J-messages and creates/updates the related MCOA entry to include the source of the J-message (as the SR), CGA to the SR's CGA (as the entry's CGA), and the state to *Joined*.

Consider the example shown in Figure 6. Assume that the mobile's MCOA is MG and after power up in the domain it connects to AR1, which then multicasts a J-message to its CAR-set (CGA1). When AR4 receives the J-message, it joins MG and creates an entry corresponding to the MCOA in *Joined* state as shown in Figure 7 (a). Later when the MN moves to AR5 it becomes the new serving router. Then AR5 sends a multicast J-message to its CAR-set (CGA5) followed by a HO message to the old serving router AR1. Since AR4 is a member of both CGA1 and CGA5, it receives both J-message from AR5 and L-message from AR1. After receiving the J-message the table entry is updated as shown in Figure 7 (b). If received after the J-message, the L-message is discarded. Thus, AR4 remains joined to MG. If received before the J-message, however, the L-message may cause AR4 to leave the MG, which interrupts packet flow to AR4 until it receives the J-message and joins the MG group. The interruption may be minimized by delaying the leave operation. In most cases the HO message delay is sufficient to minimize the interruption. A simple scheme can be employed that periodically checks the table to purge all the entries that are in the Left state and consequently prune the corresponding multicast trees.

---

[4] To account for lost L-message, or crash of the SR, a soft-state mechanism is used. SR sends periodic J-messages containing table changes (if any) and providing liveness.

| MCOA | Serving Router | CGA | State |
|------|----------------|------|--------|
| MG | AR1 | CGA1 | Joined |

(a)

| MCOA | Serving Router | CGA | State |
|------|----------------|------|--------|
| MG | AR5 | CGA5 | Joined |

(b)

**Figure 7**: Table state at AR4 (a): when MN1 is connected to AR1 (b): after MN1 moved to AR5

## V. Evaluation and Comparison

In order to evaluate the performance of M&M and compare it with other known schemes, we simulated M&M, Hawaii[21] and CIP[20] – the three routing-based mobility solutions[5]. We modified the network simulator, ns-2 [17] to incorporate M&M. We changed the implementation of mobile node and access router to add mobility detection, handover algorithm and multicast routing.

### A. Performance metrics

We used the following performance metrics to evaluate the performance of M&M and compare it to CIP and HAWAII.
- *Handoff delay* is defined as the difference between the time at which the MN received the last packet from the old access router and the first packet from the new access router.
- *Depth of packet reordering* is measured as the maximum difference in the sequence numbers of adjacent packets. This is a rough indicator of the size of the buffer needed to re-sequence the out of order packets.
- *Packet duplication* is the total number of packets duplicated in a single handoff. This is measured as the duration for which reordering occurs. Since CBR traffic is used, reordering duration gives an estimate of how many packets can be duplicated irrespective of the packet rate at the source.
- *Routing efficiency* is defined as the ratio of the number of hops between the root of the tree and the MN to the number of hops on the shortest path between the two. This gives a qualitative comparison of routing efficiency.

We did not consider packet loss as a metric for this work as it is also sensitive to factors other than handoff delay such as packet arrival rate and mobility pattern. Mobility detection need not necessarily be a part of the micro-mobility protocol as this can be better achieved with additional information from lower layers.

### B. Simulation Scenarios

To study the factors affecting the performance of the micro-mobility protocols we simulated a rich set of scenarios including both tree topologies of varying depth ranging from 3 to 6. The link bandwidths were fixed at 10Mbps for wired links with delays varied from 10ms to 5ms to 2ms for all links. Detailed 802.11 models in ns-2 were used for the wireless part with cell overlap of 30m. Beacons spacing 200ms apart are used for mobility. Prune timeout of 1s is set for the multicast protocol. The handoff mechanism for M&M, CIP and HAWAII are bi-cast, semi-soft handoff and Multi Stream Forwarding (MSF) [21] respectively. Both M&M and CIP use bi-cast technique whereby packets are bi-cast to both old and new ARs from a crossover point within the network. In contrast, HAWAII uses buffer and forward technique where the old AR buffers the packets and forwards them during route repair. Random mobility at 30m/s was the mobility pattern used for the MN. CBR traffic with packet size of 512 bytes and 10ms/packet was used. To avoid the side effects of mechanisms of other protocols (like congestion control mechanism of TCP) affecting the handoff delay and packet delivery performance, we chose CBR over UDP as opposed to FTP over TCP.

### C. Simulation results

We conducted simulations over different topologies, varying parameters like beacon timer, and link delays. Since mobility detection mechanism is not a part of the protocol, simulations were set-up such that mobility detection always happened when the MN moved from one access router to another. This was to prevent loss of packets due to failure of mobility detection.

---
[5] We have also compared our scheme to hierarchical MIP[27] and seamless handoff[31] schemes using route-based analysis. Please refer to [39] for details. As was shown in [39] M&M achieved the min handoff delay and min overhead among the three classes.

Graphs for different topologies show the same trend; hence we selected simpler graphs for the tree topology with depth 3. Figure 8 shows the topology used in the simulation.

All the graphs follow a common format. Each graph shows data for M&M, CIP and HAWAII (in that order from left to right). The x-axis shows three sets of data corresponding to link delays of 10ms, 5ms and 2ms (again from left to right) for each protocol. Path lengths from fork router to old and new access routers vary along y-axis. For example, '3,2' means path length of 3 hops from the fork router to the old access routers and 2 hops from the fork router to the new access router. The z-axis shows the performance parameters under evaluation.

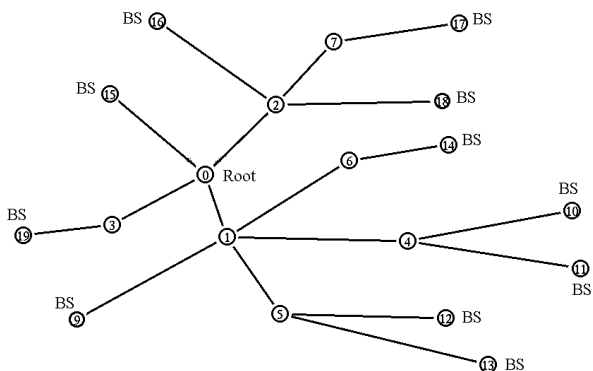

**Figure 8**: Simple tree topology

Figure 9 illustrates the handoff delays incurred by M&M, CIP and HAWAII with link delays 10, 5 and 2ms. From the graphs, we observe that the handoff delay for M&M and CIP is small as compared to that of HAWAII. Both CIP and M&M use bi-cast, which causes smooth handover with negligible handover delay. Whereas, the HAWAII using the MSF, a buffer and forward scheme consistently incurs long handoff delays.

Figures 10 shows the depth of reordered packets. We measured depth of reordering instead of the number of packets reordered because it indicates the size of buffer needed to re-sequence the out of order packets. It is obvious from the graph that the depth of reordering is small for M&M and CIP, whereas it is large for HAWAII. The out of sequence packets in M&M and CIP is dependent on the difference in the link delays from fork router to old and new access routers. The greater the difference, the greater will be the depth of reordering. In case of HAWAII the depth is large because the old access router buffers packets and then forwards it to the new access router via the crossover router. The crossover router also forwards the incoming packets to the new access router at the same time. This results in packets reaching the new access router out of order. The depth of reordering is dependent on the buffering duration and the link delays from the cross over router to the old access router. Its also important to observe the duration for which reordering of packets occur. In M&M and CIP, the reordering occurs as long as bi casting is done. However, in HAWAII, reordering duration depends on the number of packets buffered at the old access router and the link delay from the old access router to the crossover point.

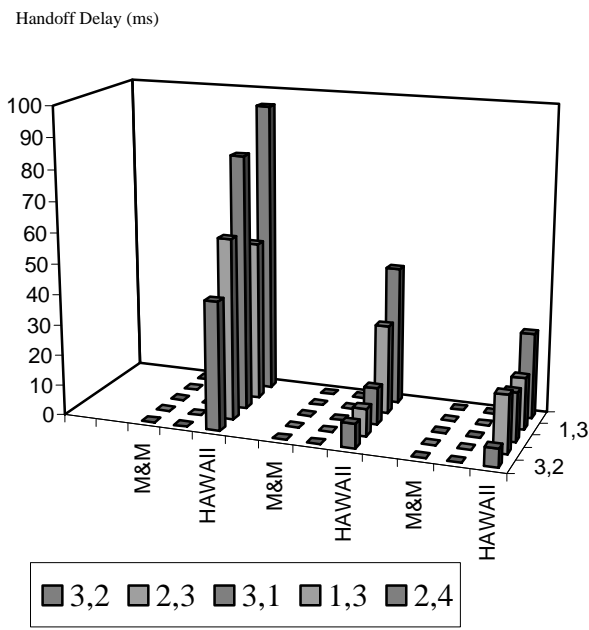

**Figure 9**: Handoff Delay

It is also important to observe the duration for which reordering of packets occur, because it indicates an estimate of the amount of packet duplication caused by a scheme. Figures 11 illustrate the duration for which reordering caused by the three schemes. In case of M&M and CIP, the reordering occurs as long as bi casting lasts causing large number of packet duplication as shown in the figure. Whereas, for HAWAII reordering duration depends on the number of packets buffered at the old access router and the link delay from the old access router to the crossover point, which shows relatively low number of duplications.

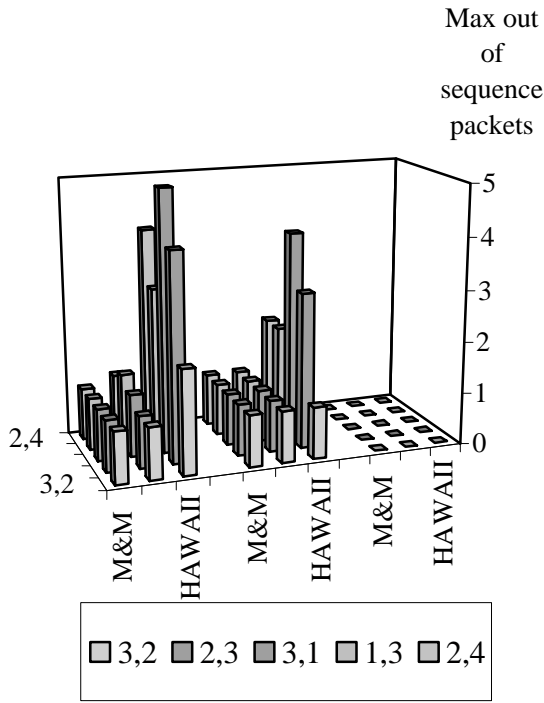

**Figure 10**: Maximum difference in sequence numbers of consecutive packets.

In case of border router (BR) acting as the root of the multicast tree the M&M uses the shortest path to route packets to the MN. This is unlike CIP, which uses the shortest path along the reverse path from the MN to the BR to route packets from the BR to the MN. Hence, it does not guarantee shortest path. However, in most cases the routing in M&M is as efficient as CIP. In case of HAWAII routing is a function of topology and node mobility, which is generally less efficient than that of M&M and CIP.

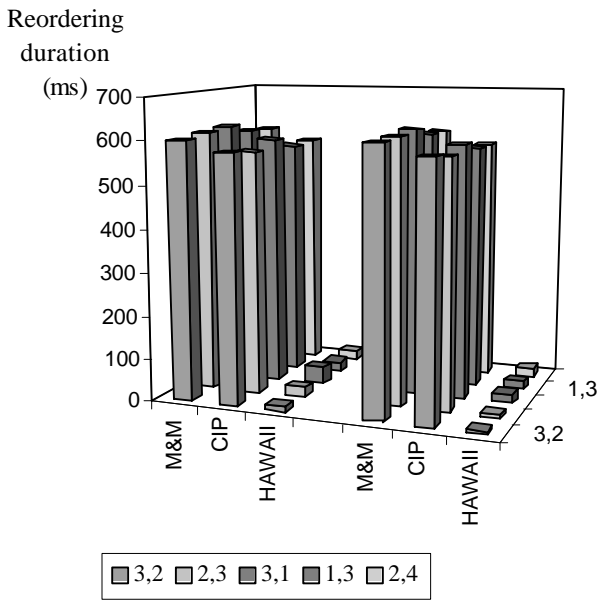

**Figure 11**: Reordering duration

Both HAWAII and CIP do not handle well the case where a domain contains multiple border routers. In particular, if packets enter the domain through one border router and leave through another border router, routing in CIP fails. The M&M relies on the underlying multicast protocol to handle multiple border routers in a domain, which is often the case. For example, mechanisms exist in PIM-SM to deliver packets to the RP irrespective of the location of the sender (BR at which the packet enters the domain). The flexibility comes at the expense of decreasing routing efficiency, because packets are first tunneled to the RP and then delivered to the MN through the multicast tree. To alleviate this situation only the BRs can be configured as candidate RP, thus ensuring that one of the BRs becomes the RP.

## VI. Related Work

Several architectures have been proposed to provide IP mobility support. In Mobile IP (MIP)[3], every mobile node (MN) is assigned a *home* address and *home agent* (HA) in its home subnet. When the MN moves to another *foreign* subnet, it acquires a care-of-address (COA) through a foreign agent (FA). The MN informs the HA of its COA through a *registration* process. Packets destined to the MN are sent first to the HA, then are *tunneled* to the MN. This is known as *triangle routing*, a major drawback of MIP. *Route optimization* [4] attempts to avoid triangle routing by sending *binding updates*, containing the current COA of the MN to the correspondent node (CN). However, communication overhead during handover renders this scheme unsuitable for *micro* mobility. In[16] end-to-end IP mobility is proposed, based on dynamic DNS updates. When MN moves, it obtains a new IP-address and updates the DNS mapping for its host name. This incurs handover latency due to DNS update delays and is not suitable for delay-bounded applications. Also, the scheme is not transparent to the transport protocol that is aware of the mobility.

In[10] the HA tunnels packets using a pre-arranged multicast group address. The access router, to which the MN is currently connected, joins the group to get data packets over the multicast tree. This approach suffers from the triangle routing problem; packets are sent to HA first and then to MN. Multicast-based mobility is proposed in [1] and [8]. Each MN is assigned only a unique multicast address. Packets sent to the MN are destined to that multicast address and flow down the multicast distribution tree to the MN. The CN tunnels the packets using the multicast address. This approach avoids triangle routing, in addition to reducing handover latency and packet loss. The study in[1] quantifies the superiority of handover performance for multicast-based mobility over Mobile IP protocols. These schemes, however, suffer from several serious practical issues, including scalability of multicast state, address allocation and dependency on inter-domain multicast. We address these issues in our work.

Several approaches have been proposed for *micro* mobility[18]. The general approaches include mobile-specific routing, hierarchical approaches and seamless handover. Mobile-specific route approaches include cellular IP[20] and Hawaii[21]. A domain-gateway registers its address with the HA (this has similarities to our proxy-based approach) and forwards the packets to the MN. The MN's home address is used within the domain. These approaches need special signaling to update mobile-specific routes and require changes in packet forwarding and unicast routing in all the routers. In cellular IP[20], signaling is data-triggered to create paths by having routers snoop on the data packets. Hawaii[21] proposes a separate routing protocol and requires explicit signaling from the mobiles. In a way, these approaches attempt to create a distribution tree using extra routing entries for the mobile, similar to what multicast does. Our approach builds upon existing multicast mechanisms as opposed to re-creating them. Approaches based on seamless handover between old and new access routers, involve fairly complex signaling, buffering and synchronization procedures. Router-assisted smooth handoff in MIP[5], and edge mobility[22] belong to this category. Fast Handover [31] introduces fast tunnel set-up between $AR_{old}$ and $AR_{new}$ as soon as the layer 2 handoff is detected. The tunnel avoids packet losses caused by path set-up delay inside the mobility domain. In a way it is complementary to our multicast-based routing inside the mobility domain. Unlike fast handover, however, our m-subnet idea considers the edge of the network as a single subnet and allows mobile node to carry RCOA and MCOA across ARs, which reduces the handover latency. Approaches using a hierarchy employ a gateway per-domain and need to keep a location database to map identifiers into locations. This mapping suffers from scalability and robustness problems as was noted earlier in this paper. In[12] a hierarchy of foreign agents is created at the local, administrative domain and global levels. In[19] a multi-level hierarchy is used in which packets from the HA arrive at a root FA where they are tunneled to a lower level FA and then to the MN. Hierarchical MIP[27] builds a network of tunnels (overlay network) between FAs. [23][29] also use a notion of mobility agent for localized handoff within a domain. We have shown in [39] that our multicast-based intra-domain mobility scheme outperforms seamless handover and hierarchical approaches and is simpler. This result is consistent with the comparison of routing-based (HAWAII and CIP) and tunneling-based (Hierarchical Mobile IP) schemes reported in [40]. It is shown that Hierarchical Mobile IP performs either equally well or inferior to the routing-based schemes, because it does not take advantage of the proximity of crossover router to serving AR.

## VII. Concluding Remarks

We have presented a novel approach to IP micro-mobility using intra-domain multicast-based mobility. Our approach solves major challenging problems facing the deployment of multicast-based mobility. In terms of multicast state scalability we note that the multicast state growth is O($G$) for the architecture presented in this study, as opposed to O($S$x$G$) in[1][8]. Our novel algorithmic mapping scheme from unicast to multicast address ensures collision-free assignment by providing unique and consistent mapping throughout the network. This solves the address allocation problem and provides robustness and per-domain privacy as multicast packets are not forwarded out of the domain. In addition, we present a new proactive path setup scheme to improve handover performance. Our extensive simulations show that:

- There is a significant difference in handoff delay and packet reordering performance between protocols using different types of handoff schemes. For example, M&M and CIP use bi-cast while HAWAII use buffer and forwarding.
- In most cases the M&M and CIP show comparable routing efficiency and handoff performance because both use shortest path routing as opposed to HAWAII. Routing packets on the path that is not the shortest path from the root of the tree to the MN not only increases end-to-end delay, but also wastes bandwidth and creates extra mobile specific routing entries.
- Bi casting:

- Masks handoff delays (handoff delay is zero)
- Produces large number of duplicate packets
- Shows small reordering depth depending on the difference in the path lengths from the fork router to the old and new access routers
- Buffering and forwarding
  - Incurs longer handoff delays
  - May produce large reordering depth

For proactive handover M&M performs as well as CIP, and it handles the case of multiple BR in a domain better than others. The M&M scheme is expected to outperform CIP in reactive handover because of its proactive path setup capability. It uses multicast routing protocol, e.g. PIM-SM, which is more reliable with readily available robust implementation and people having more experienced managing it. All these factors facilitate the deployment of M&M in wireless service provider domain. Furthermore, it naturally supports efficient multicasting to MNs.

In future, we plan to extend our simulator for simulating reactive handover scenarios. We also would like to develop M&M support for efficient mobile-mobile communication.